\documentstyle[12pt]{article}
\input epsf

\def\rulerheight{0.5pt}

\def\U1{$U(1)$}
\def\SU5{$SU(5)$}
\def\SO10{$SO(10)$}
\def\422{$SU(4)\otimes SU(2)_L \otimes SU(2)_R$}

\def\diag.{\hbox{diag.}}
\def\muegamma{\hbox{$\mu\to e+\gamma$\ }}
\def\taumugamma{\hbox{$\tau\to\mu+\gamma$\ }}

\def\M_U{\hbox{$M_U$}\ }
\def\M_P{\hbox{$M_P$}\ }

\def\tanb{\hbox{$\tan \beta$}}

\def\MSSM+N{\hbox{MSSM+$\nu$}}
\def\etal{{\it et al.}}
\def\bigsim{{\>{\buildrel {\scriptstyle  >} \over {\scriptstyle \sim} }\>}}
\def\smlsim{{\>{\buildrel {\scriptstyle <} \over {\scriptstyle \sim} }\>}}

\def\ibid{{\it ibid.}}

\newcommand\beq{\begin{equation}}
\newcommand\eeq{\end{equation}}
\newcommand\bea{\begin{eqnarray}}
\newcommand\eea{\end{eqnarray}}
\newcommand\ba{\begin{array}}
\newcommand\ea{\end{array}}

\begin{document}
\baselineskip 24pt
\newcommand{\sheptitle}
{Muon Anomalous Magnetic Moment and $\tau \rightarrow \mu \gamma $ \\
 in a 
 Realistic String-Inspired Model of Neutrino Masses}

\newcommand{\shepauthor}
{T. Bla\v{z}ek$^*$ and S. F. King}

\newcommand{\shepaddress}
{Department of Physics and Astronomy, University of Southampton \\
        Southampton, SO17 1BJ, U.K}

\newcommand{\shepabstract}
{We discuss the lepton sector of
a realistic string-inspired model
 based on the Pati-Salam $SU(4)\times SU(2)_L \times SU(2)_R$ 
 gauge group supplemented by a $U(1)$ family symmetry. 
 The model involves third family Yukawa unification, predicts
 large $\tan \beta\sim 50$, and describes all fermion masses and
mixing angles, including approximate bi-maximal mixing in the neutrino
sector. Atmospheric neutrino mixing is achieved via a large
 23 entry in the neutrino Yukawa matrix which can have important
phenomenological effects. 
We find that the recent
 BNL result on the muon ($g-2$) can be easily accommodated
in a large portion of the SUSY parameter space of this model.
Over this region of parameter space the model predicts
a CP-even Higgs mass 
near 115 GeV, and a rate for \taumugamma which is close
to its current experimental limit.
%
%
}

\begin{titlepage}
\begin{flushright}
hep-ph/0105005
\end{flushright}
\begin{center}
{\large{\bf \sheptitle}}
\\ \shepauthor \\ \mbox{} \\ {\it \shepaddress} \\ 
{\bf Abstract} \bigskip \end{center} \setcounter{page}{0}
\shepabstract
\begin{flushleft}
\today
\end{flushleft}

\vskip 0.1in
\noindent
$^*${\footnotesize On leave of absence from 
the Dept. of Theoretical Physics, Comenius Univ., Bratislava, Slovakia}

\end{titlepage}

\newpage

Recently the BNL E821 Muon g-2 Collaboration has reported a precise 
measurement of the muon anomalous magnetic moment \cite{muon}
$a_{\mu}(exp)=(g-2)/2$,
\beq
a_{\mu}(exp)=(11,659,202 \pm 15)\times 10^{-10}.
\eeq
When combined with the other four most recent measurements
the world average of $a_\mu$ is now higher than 
the Standard Model (SM) prediction,
\beq
a_{\mu}(SM)=(11,659,160 \pm 7)\times 10^{-10}
\eeq
by $(43\pm16)\times 10^{-10}$ which corresponds to a discrepancy of 
$2.6\sigma$.
It is well known that Supersymmetry (SUSY) gives an additional contribution
to $a_{\mu}(SM)$ which is dominated by the chargino exchange diagram and
approximately given by
\beq
\Delta a_{\mu}(SUSY)\sim \frac{\alpha_2}{4\pi}
\left(\frac{\mu M_2m_{\mu}^2}{M_{SUSY}^4}\right) \tan \beta 
\eeq
where $\alpha_2$ is the $SU(2)$ gauge coupling, 
$\mu$ is the SUSY Higgs mass parameter, 
$M_2$ is $SU(2)$ gaugino mass,
$m_{\mu}$ is the muon mass,
$M_{SUSY}$ represents the 
heaviest sparticle mass in the loop,
and $\tan \beta$ is the ratio of Higgs vacuum expectation values (VEVs).
Note that the sign of $\Delta a_{\mu}(SUSY)$ 
depends on the sign of $\mu$ (relative to $M_2$).

Well before the experimental result from BNL was published,
it was realised that the additional SUSY contribution $\Delta a_{\mu}(SUSY)$ 
could be of the correct order of magnitude to be observed
by E821 providing that $\tan \beta$ is
sufficiently large, and the relevant superpartner masses $M_{SUSY}$ are
not too large \cite{old_amu_SUSY}.
Since the reported result, there has been a blizzard of theoretical
papers, showing how the result may be accomodated within
SUSY in detail and for various models \cite{blizzard}.
The general concensus of these recent studies is that
numerically the additional SUSY contribution is sufficient
to account for the discrepancy between the SM value
and the experimental value, providing that $\tan \beta \bigsim 10$ and
$M_{SUSY}\smlsim 500$ GeV, and of course that the sign of $\mu$ is positive.

Large $\tan \beta$ is also required in order to
have a Higgs boson of mass 115 GeV \cite{115}
(where the LEP signal has a significance of $2.9\sigma$)
and it is encouraging that both signals point in the same
direction of large $\tan \beta$. It is even more encouraging
that some well motivated unified models have long predicted that
$\tan \beta$ is large. In particular models based on the gauge groups
\SO10 or the Pati-Salam group \422 predict Yukawa unification
which in turn implies $\tanb \sim 50$ 
\cite{KiSh,KiOl2}. Is experiment 
giving us a hint that Nature favours one of these Yukawa unification
models which predict large \tanb? 

There is a further piece of experimental evidence in favour of 
these models, namely that they both contain gauged $SU(2)_R$
symmetry and hence they both predict three right-handed neutrinos
and hence non-zero neutrino masses. Thus in these models neutrino 
masses are compulsory, and not optional as in $SU(5)$ for
example. SuperKamiokande evidence for atmospheric neutrino oscillations
\cite{SKamiokandeColl} has taught us that
neutrino masses are non-zero and furthermore that the 23 mixing angle
is almost maximal. The evidence for solar neutrino oscillations
is almost as strong, although the conclusions are more ambiguous
\cite{GoPe}. A minimal interpretation of the atmospheric and solar
data is to have a three neutrino hierarchy. A simple
and natural interpretation of the data is 
single right-handed neutrino dominance (SRHND) \cite{SRHND}. 
In a large class of models, including those with SRHND, 
the large atmospheric mixing angle is due to large and
equal couplings in the 23 and 33 entries of the Dirac neutrino
Yukawa matrix (in the LR basis)
\beq
Y_{\nu}
\sim 
\left( \begin{array}{ccc}
0 & 0 & 0    \\
0 & 0 & 1    \\
0 & 0 & 1   
\end{array}
\right)
\label{Yuk}
\eeq
corresponding to the dominant third right-handed neutrino coupling
equally to the second and third lepton doublets. 
The see-saw mechanism yields a physical neutrino with a mass
about $5\times 10^{-2}$ eV consistent with the SuperKamiokande 
observation providing the third right-handed neutrino mass is
$M_{R3} \approx 3\times 10^{14}$ GeV \cite{KiOl2}.

The large off-diagonal Yukawa coupling in Eq.\ref{Yuk} will have
an important effect on the 23 block of the slepton doublet soft mass squared
matrix $m_L^2$, when the renormalisation group equations (RGEs)
are run down from $M_{GUT}$ to the mass scale of the
third right-handed neutrino $M_{R3}$. In order to see this it is
instructive to examine the RGEs for $m_L^2$,
\bea
\frac{d m_L^2}{dt} & = & \left(\frac{d m_L^2}{dt}\right)_{Y_{\nu}=0}
\nonumber \\
& - & \frac{1}{32\pi^2}\left[
Y_{\nu}Y_{\nu}^\dagger m_L^2+m_L^2Y_{\nu}Y_{\nu}^\dagger
+2Y_{\nu}m_N^2Y_{\nu}^\dagger+2(m_{H_u}^2)Y_{\nu}Y_{\nu}^\dagger
+2\tilde{A}_{\nu}\tilde{A}_{\nu}^\dagger
\right]
\eea
where $m_N^2$, $m_{H_u}^2$ are the soft mass squareds of the
right-handed sneutrinos and up-type Higgs doublet, 
$\tilde{A}_{\nu}$ is the soft trilinear mass parameter associated 
with the neutrino Yukawa coupling, and $t=\ln (M_{GUT}^2/\mu^2)$,
where $\mu $ is the $\bar{MS}$ scale.
The first term on the right-hand side represents 
terms which do not depend on the neutrino Yukawa coupling.
Assuming universal soft parameters at $M_{GUT}$,
$m_L^2(0)=m_N^2(0)=m_0^2I$, where $I$ is the
unit matrix, and $\tilde{A}_{\nu}(0)=A Y_{\nu}$, we have
\beq
\frac{d m_L^2}{dt}  =  \left(\frac{d m_L^2}{dt}\right)_{Y_{\nu}=0}
 -  \frac{(3m_0^2+A^2)}{16\pi^2}\left[ Y_{\nu}Y_{\nu}^\dagger \right]
\eeq
where in the basis in which the charged lepton Yukawa couplings
are diagonal, the first term on the right-hand side
is diagonal. In running the RGEs between $M_{GUT}$ and $M_{R3}$ the neutrino
Yukawa couplings lead to an approximate contribution to the
slepton mass squared matrix of
\beq
\delta m_L^2 \approx -\frac{1}{16\pi^2}
\ln \left(\frac{M_{GUT}^2}{M_{R3}^2}\right)
(3m_0^2+A^2)
\left[ Y_{\nu}Y_{\nu}^\dagger \right]
\approx -0.1(3m_0^2+A^2)\left[ Y_{\nu}Y_{\nu}^\dagger \right]
\label{delta}
\eeq

Using the SRHND form of the neutrino Yukawa matrix
in Eq.\ref{Yuk} we find 
\beq
Y_{\nu}Y_{\nu}^\dagger
\sim 
\left( \begin{array}{ccc}
0 & 0 & 0    \\
0 & 1 & 1    \\
0 & 1 & 1   
\end{array}
\right)
\label{Yuksq}
\eeq
and according to Eq.\ref{delta} the large neutrino Yukawa coupling
in the 23 position will imply an off-diagonal
23 flavour violation in the slepton mass squared matrix which
will be of order 5-10\% of the diagonal soft mass squareds,
and will be observable in the lepton flavour violating (LFV)
process \taumugamma.
In addition the 22 entry of the slepton mass squared matrix will
receive a 5-10\% correction which again is due to the large 23 neutrino
Yukawa coupling, and is much larger than the usual correction
due to the diagonal muon Yukawa coupling which is very small.
The large 22 entry in Eq.\ref{Yuksq} will thus 
give a significant correction to the
relation between the GUT scale soft mass parameters and
the muon (g-2) estimates. The main purpose of the present paper
is to explore these observable effects in the framework of
a particular model which predicts Yukawa unification, 
and hence large \tanb, namely the string-inspired Pati-Salam model
based on the gauge group \422 \cite{PaSa}.

For completeness we briefly review the string-inspired Pati-Salam model.
As in $SO(10)$ the presence of the gauged $SU(2)_R$ predicts the
existence of three right-handed neutrinos.
However, unlike $SO(10)$, there is no Higgs doublet-triplet
splitting problem since 
both Higgs doublets are unified into
a single multiplet $h$.
Heavy Higgs $H,\bar{H}$ are introduced 
in order to break the symmetry.
The model leads to third family
Yukawa unification, as in minimal $SO(10)$, and the phenomenology
of this was recently discusssed \cite{KiOl2}.
Although the Pati-Salam gauge group is not unified at the
field theory level, it readily emerges from string constructions
either in the perturbative fermionic constructions \cite{AnLe},
or in the more recent type I string constructions \cite{ShTy},
unlike $SO(10)$ which typically requires large Higgs representations
which do not arise from the simplest string constructions.

The Pati-Salam gauge group \cite{PaSa},
supplemented by a $U(1)$ family symmetry, is
\begin{equation}
SU(4) \otimes SU(2)_L \otimes SU(2)_R\otimes U(1)
\end{equation}
with left (L) and right (R) handed 
fermions transforming as $F_L\sim (4,2,1)$
and $F_R\sim ({4},1,2)$
in the superfield multiplets
\begin{equation}
{F^i_{L,R}}=
\left(\begin{array}{cccc} 
{u}  & {u} & {u} & {\nu} \\  
{d} & {d} & {d} & {e^-}     
\end{array} \right)_{L,R}^i     
\end{equation}
The Higgs $h$ contains the two MSSM Higgs doublets
and transforms as $h\sim (1,2,2)$
\begin{equation}
h=
\left(\begin{array}{cc}
{h_1}^0 & {h_2}^+ \\   
{h_1}^- & {h_2}^0      
\end{array} \right) 
\end{equation}
The Higgs $H,\bar{H}$ transform as 
$H\sim (4,1,2)$, $\bar{H}\sim (\bar{4},1,2)$ 
and develop VEVs which break the Pati-Salam group,
while $\theta, \bar{\theta}$ are Pati-Salam singlets
and develop VEVs which break the $U(1)$ family symmetry.
\begin{equation}
{H},\bar{H} =
\left(\begin{array}{cccc}
{u_H} & { u_H} & { u_H} & { \nu_H} \\
{ d_H} & { d_H} & { d_H} & { e_H^-}   
\end{array} \right),\cdots
\end{equation}
We assume for convenience that all symmetry breaking
scales are at the GUT scale,
\begin{equation}
<H>=<\bar{H}>=<{ \nu_H}>\sim M \sim 10^{16}GeV
\end{equation}
\begin{equation}
<\theta>=<\bar{\theta}>\sim M \sim 10^{16}GeV
\end{equation}

The fermion mass operators (responsible for Yukawa
matrices $Y_u$,$Y_d$,$Y_e$,$Y_{\nu}$) are \cite{Ops}:
\begin{equation}
(F^i_L \bar{F}^j_R )h\left(\frac{H\bar{H}}{M^2}\right)^n
\left(\frac{\theta}{M}\right)^p
\end{equation}
The third family is assumed to have zero $U(1)$ charge,
and the 33 operator is assumed to be the renormalisable operator with $n=p=0$
leading to Yukawa unification. 
The remaining operators have $n>0$ with varying group contractions
involving $H\bar{H}$ leading to different Clebsch factors. The latter
are responsible for vertical mass splittings within a generation.
The mass splittings between different generations are
described by operators with $p>0$ arising from different
$U(1)$ charge assignments to the different families. 
The Majorana mass operators (responsible for $M_{RR}$) are \cite{Ops}:
\begin{equation}
(\bar{F}^i_R\bar{F}^j_R )\left(\frac{HH}{M^2}\right)
\left(\frac{H\bar{H}}{M^2}\right) ^m
\left(\frac{\theta}{M}\right) ^q.
\end{equation}

We recently discussed \cite{KO} neutrino masses
and mixing angles in the above string-inspired Pati-Salam model
supplemented by a $U(1)$ flavour symmetry. 
We used the SRHND mechanism, which
may be implemented in the 422 model by having a 23 operator with
$p=0$ and $n=1$ where the Clebsch is non-zero in the neutrino
direction, but zero for charged fermions. This results 
in a natural explanation for atmospheric neutrinos
via a hierarchical mass spectrum. We specifically
focused on the LMA MSW solution since this is slightly preferred by the
most recent fits, and assuming this
a particular model of high energy
Yukawa matrices which gave a good fit to all quark and lepton masses
and mixing angles was discussed \cite{KO}. 
The numerical values
of the high energy Yukawa matrices 
in this example are reproduced
in Table I.
 To study lepton flavour violation focusing on
 the effects of the large off-diagonal 23 entry 
 in $Y_\nu$, in this study we have further suppressed 
 the tiny entries ${Y_e}_{12}$,
 ${Y_e}_{13}$, and ${Y_\nu}_{13}$ compared to the values quoted in 
 \cite{KO}. 
 Note that with the suppression above 
 the branching ratio $BR($\muegamma$)$ stays well below the 
 experimental limit, without substantially changing the predictions of
 fermion masses and mixing angles. This
 demonstrates that this channel is more
model dependent than \taumugamma which is our main focus in this paper.

The neutrino Yukawa matrix in Table I has a similar structure to that 
discussed in Eq.\ref{Yuk} and has large approximately equal
23 and 33 elements. Thus the Yukawa matrices in Table I are examples
of the effect that leads to 5-10\% corrections to the 23 block
of the slepton mass squared matrix $m_L$ that we discussed previously.
We now turn to a numerical discussion of these effects.

\vbox{
\vbox{
\begin{center}
\begin{tabular}{ccc}
\noalign{\medskip}
\noalign{\hrule height\rulerheight}
\noalign{\smallskip}
\noalign{\hrule height\rulerheight}
\noalign{\medskip}
$
Y_u(M_X) $ & $=$ & $\left(\matrix{
\phantom{-}
7.034\times 10^{-6} &
\phantom{-}
4.079\times 10^{-4} &
\phantom{-}
4.324\times 10^{-3} \cr
\phantom{-}
3.991\times 10^{-5} &
\phantom{-}
1.466\times 10^{-3} &
\phantom{-}
0.000 \cr
\phantom{-}
3.528\times 10^{-5} &
-
3.748\times 10^{-3} &
\phantom{-}
0.677 }\right) 
$ \\
\noalign{\smallskip}
$
Y_d(M_X)$ & $=$ & $\left(\matrix{
-
2.331\times 10^{-4} &
-
4.079\times 10^{-4} &
\phantom{-}
8.648\times 10^{-3} \cr
\phantom{-}
4.609\times 10^{-4} &
-
8.827\times 10^{-3} &
\phantom{-}
2.157\times 10^{-2} \cr
-
8.246\times 10^{-4} &
\phantom{-}
1.506\times 10^{-2} &
\phantom{-}
0.677 }\right)
$ \\
\noalign{\smallskip}
$
Y_e(M_X)$ & $=$ & $\left(\matrix{
-
1.748\times 10^{-4} &
\phantom{-}
3.884\times 10^{-5} &
\phantom{-}
 8.574\times 10^{-4} \cr
\phantom{-}
9.219\times 10^{-4} &
\phantom{-}
 3.015\times 10^{-2} &
-
 6.472\times 10^{-2} \cr
-
6.184\times 10^{-4} &
\phantom{-}
 1.501\times 10^{-2} &
\phantom{-}
0.677}\right)
$ \\
\noalign{\smallskip}
$
Y_\nu(M_X)$ & $=$ & $\left(\matrix{
\phantom{-}
 7.034\times 10^{-6} &
\phantom{-}
 2.401\times 10^{-3} &
\phantom{-}
 7.710\times 10^{-4} \cr
\phantom{-}
 2.993\times 10^{-5} &
\phantom{-}
 2.932\times 10^{-3} &
\phantom{-}
 0.440 \cr
\phantom{-}
 3.528\times 10^{-5} &
-
 2.811\times 10^{-3} &
\phantom{-}
 0.677}\right)
$ \\
\noalign{\smallskip}
$
M_{RR}(M_X)$ & $=$ & $\left(\matrix{
\phantom{-}
 3.991 \times 10^{ 8}\phantom{^1} &
\phantom{-}
 5.652 \times 10^{ 9}\phantom{^1} &
\phantom{-}
 1.040 \times 10^{11} \cr
\phantom{-}
 5.652 \times 10^{ 9}\phantom{^1} &
\phantom{-}
 1.706 \times 10^{11} &
\phantom{-}
 1.866 \times 10^{12} \cr
\phantom{-}
 1.040 \times 10^{11} &
\phantom{-}
 1.866 \times 10^{12} &
\phantom{-}
 3.090 \times 10^{14} \cr
}\right)
$ \\
\noalign{\medskip}
\noalign{\hrule height\rulerheight}
\noalign{\smallskip}
\noalign{\hrule height\rulerheight}
\end{tabular}
\end{center}
{\footnotesize Table I. Yukawa matrices at $M_{GUT}$ (from
ref.\cite{KO}) where the matrix elements of $M_{RR}$ are in GeV.}
}}

In our numerical analysis we have adopted a complete top-down approach
\cite{BCRW}. 
At the GUT scale we kept $1/\alpha_{GUT}=24.5223$, 
                         $       M_{GUT}=3.0278\times 10^{16}$GeV,
$\epsilon_3\equiv (\alpha_3(M_{GUT})-\alpha_{GUT})/\alpha_{GUT} =
-4.0568\%$, and the matrices in table I as fixed.
Here $\alpha_{GUT} = \alpha_{2L} = \alpha_1$, and 
     $\alpha_3     = \alpha_4$.
For simplicity, the soft scalar masses of the MSSM superfields 
were introduced at the same scale. Including the $D$ terms from the
breaking of the Pati-Salam gauge group they read \cite{KiOl2}
\begin{equation}
\begin{array}{lcl}
                   \mbox{\rule[-0.40cm]{0mm}{0.5cm}}
   m_Q^2     &=& m_{F_L}^2 +  g_4^2\,D^2 \\
                   \mbox{\rule[-0.40cm]{0mm}{0.5cm}}
   m_{u_R}^2 &=& m_{F_R}^2 - (g_4^2-2g_{2R}^2) \, D^2 \\
                   \mbox{\rule[-0.40cm]{0mm}{0.5cm}}
   m_{d_R}^2 &=& m_{F_R}^2 - (g_4^2+2g_{2R}^2) \, D^2 \\
                   \mbox{\rule[-0.40cm]{0mm}{0.5cm}}
   m_L^2     &=& m_{F_L}^2 - 3g_4^2\,D^2 \\
                   \mbox{\rule[-0.40cm]{0mm}{0.5cm}}
   m_{e_R}^2 &=& m_{F_R}^2 +(3g_4^2-2g_{2R}^2) \, D^2 \\
                   \mbox{\rule[-0.40cm]{0mm}{0.5cm}}
 m_{\nu_R}^2 &=& m_{F_R}^2 +(3g_4^2+2g_{2R}^2) \, D^2 \\
                   \mbox{\rule[-0.40cm]{0mm}{0.5cm}}
   m_{H_u}^2 &=& m_{h  }^2 -        2g_{2R}^2  \, D^2 \\
                   \mbox{\rule[-0.40cm]{0mm}{0.5cm}}
   m_{H_d}^2 &=& m_{h  }^2 +        2g_{2R}^2  \, D^2.
\end{array}
\label{eq:D}
\end{equation}
In the numerical analysis we kept the equality between the two 
soft SUSY breaking scalar masses $m_{F_L}=m_{F_R}\equiv m_{F}$.
Two-loop RGEs for the dimensionless couplings and 
one-loop RGEs for the dimensionful couplings were used to 
run all couplings down to the scale $M_{3R}$ where the heaviest
right-handed neutrino decoupled from the RGEs. Similar steps
were taken for the lighter $M_{2R}$ and $M_{1R}$ scales, 
and finally with all three right-handed neutrinos decoupled
the solutions for the MSSM couplings were computed at the $Z$ scale.
$m_h$ and $D$ in Eqs.\ref{eq:D} 
were varied to optimize radiative electroweak symmetry breaking (REWSB), 
which was checked at one loop
following the effective potential method in \cite{EPM}.
As $\tan\beta$ determines the Higgs bilinear parameter $B\mu$,
there is a redundancy in our procedure since two input parameters, 
$m_h$ and $D$, determine one condition for the Higgs VEV of $246\,$GeV.
This freedom was removed by favouring solutions with low 
$CP$ odd Higgs mass $m_{A^0}$ as a result of the observation
that values of $m_{A^0}$ at the upper end of the range
allowed by REWSB at a given $(m_F,M_{1/2})$ point 
are correlated, through the choice of the $D$, with low
values for the stau mass which then in turn push the branching ratio 
$BR($\taumugamma$)$ above the experimental limit. For this reason
we introduced a mild penalty $\chi^2$ into our analysis to favour
REWSB solutions with low values of $m_{A^0}$.
This top-down approach enabled us to control the $\mu$ parameter
as well as $\tan\beta$. We explored regions with $\mu$ low 
($\mu = 120$GeV) and high ($\mu = 300$GeV) \footnote
{
For $\tan\beta$ as large as 50,   $\mu\gg 300$GeV 
leads to too large SUSY threshold corrections to the
masses of the third generation fermions $\tau$ and $b$
unless the sparticles in the loop have masses well above 
the $1$ TeV region.
\cite{large_mb,BCRW}
}.
As a reference point we kept $\tan\beta=50$, and
the universal trilinear coupling $A=0$.
An experimental lower bound on each sparticle mass was imposed. 
In particular,
the most constraining are: the LEP limits on the charged SUSY masses
($m_{\tilde{\chi}^\pm},m_{\tilde{\tau}}>105$GeV), the CDF limit
on the mass of the $CP$ odd Higgs state 
($m_{A^0}>105$-$110\,$GeV, valid for $\tan\beta\approx 50$) \cite{Tevatron}, 
and the requirement that the lightest SUSY particle should be neutral.
\footnote
{
 Note that in this study we are primarily concerned with the
 lepton sector of the model and the effects of the large 
 23 element of the $Y_\nu$ in Eq.4. For this reason we drop
 two important constraints in the quark sector from the 
 analysis. In particular, we do not consider the constraints 
 imposed by the $BR(b\rightarrow s\gamma)$ and accept the
 $b$ quark mass heavier than the value in \cite{PDG}
 by about 15\%. We assume that the complete theory at the high
 energy scale will induce additional corrections to the quark yukawa
 couplings possibly through a set of higher dimensional operators
 of the form (15) modifying the quark input parameters in table I.
}

\begin{figure}[p]
\epsfysize=6.5truein
\epsffile{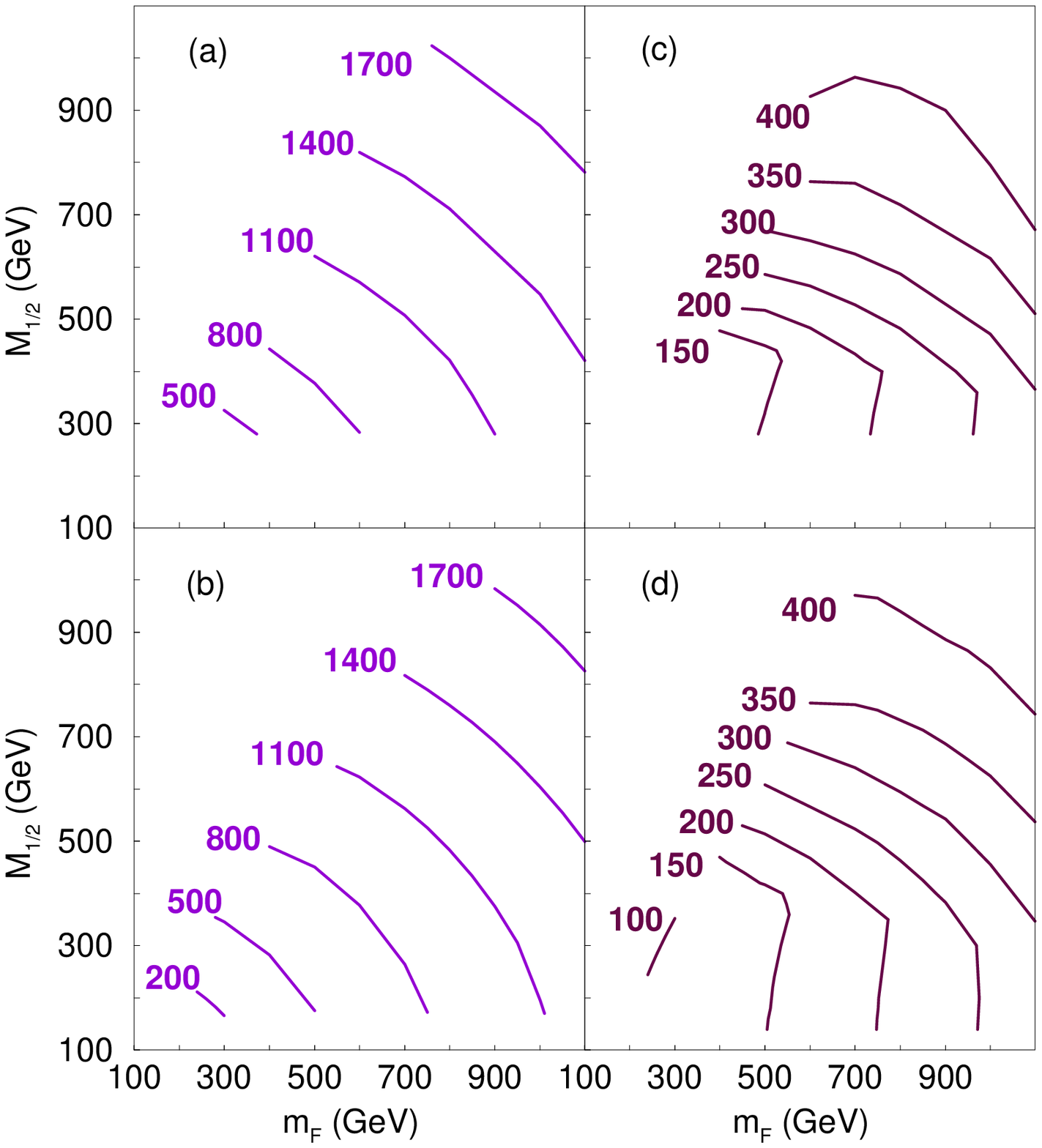}
\caption{
Contour lines of GUT scale parameters $m_h$ and $D$
determined by the condition of the radiative electroweak 
symmetry breaking, for two different values of $\mu$.
(a) $m_h$, for $\mu=120$GeV. $\;$
(b) $m_h$, for $\mu=300$GeV. $\;$
(c) $D$,   for $\mu=120$GeV. $\;$
(d) $D$,   for $\mu=300$GeV. $\;$ 
Values in the plots are in GeV. In all plots
$\tan\beta=50$, $A=0$, and $m_F=m_{F_L}=m_{F_R}$.
}
\label{f:mh_and_D}
\end{figure}

The results are presented as plots in the $(m_F,M_{1/2})$ plane.
In figure \ref{f:mh_and_D} we show the best fit values for
the quantities at the GUT scale which were varied to obtain the 
electroweak symmetry breaking.
As explained in the previous paragraph these values are not unique,
but preferred. We note that, 
clearly, the $D$ terms in Eqs.\ref{eq:D} are just a fraction
of the scalar mass $m_F$ while the scalar higgs mass parameter $m_h$ is
generally found to be greater than $m_F$.
The sharp turns in the contour
lines of constant $D$ below $M_{1/2}\approx400 GeV$ result from 
the pseudoscalar Higgs mass $m_{A^0}$ reaching the 
experimental lower bound, as demonstrated on plots (c) and (d)
in figure 2.
The parameters $m_h$ and $D$ can still adapt to this change 
for $M_{1/2} < 400 GeV$. 
The allowed 
$(m_F,M_{1/2})$ region is finally bounded from below 
because of the too low chargino mass. This bound is 
at $M_{1/2}\approx 280$GeV for $\mu=120$GeV, and 
$M_{1/2}\approx 140$GeV for $\mu=300$GeV.
The region to the left of the contour lines is disallowed 
due to the stau lighter than any neutral SUSY particle.

\begin{figure}[p]
\epsfysize=6.5truein
\epsffile{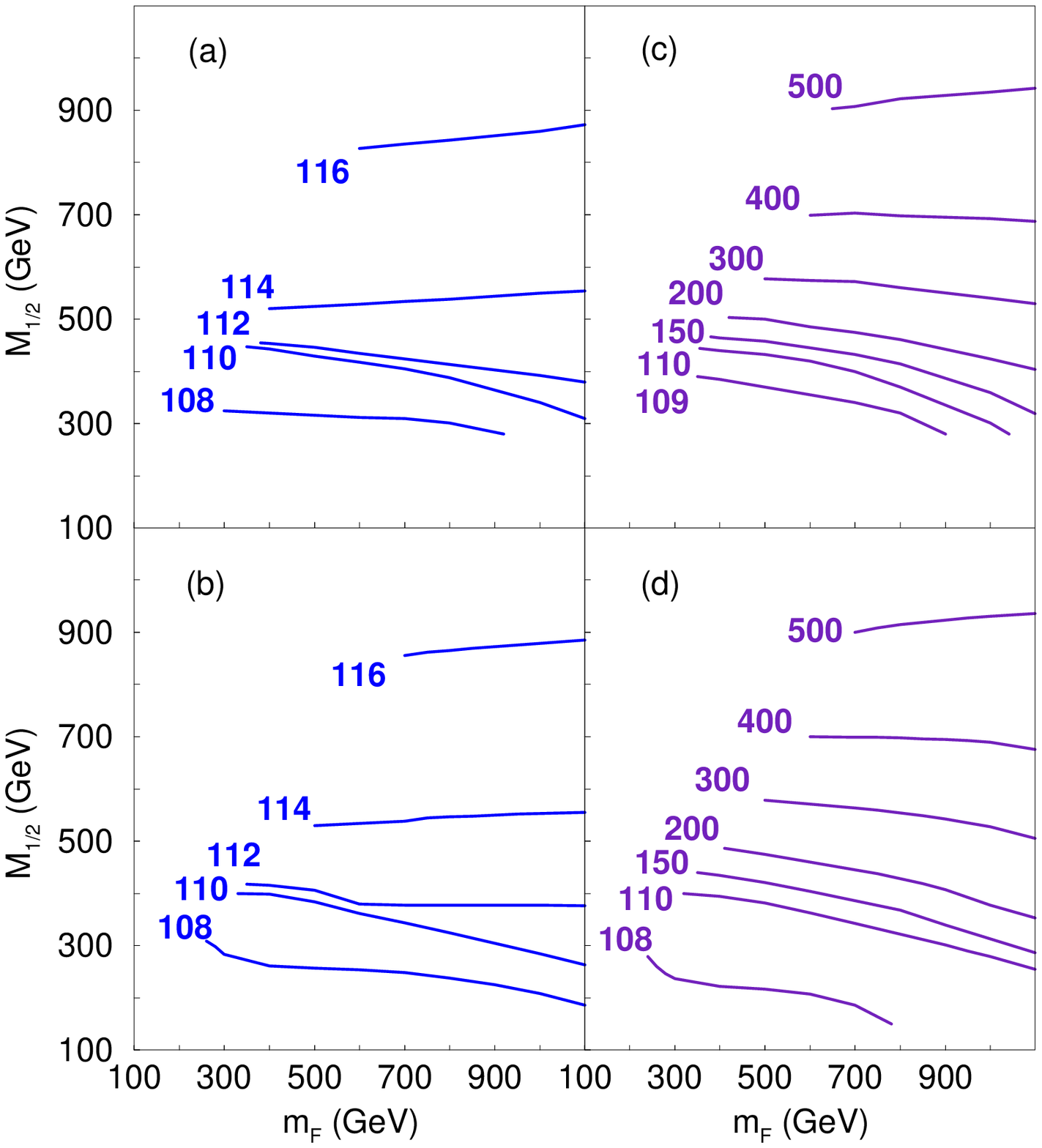}
\caption{
Contour lines of the light $CP$ even Higgs mass 
$m_{h^0}$ and pseudoscalar Higgs mass $m_{A^0}$, 
for two different values of $\mu$.
(a) $m_{h^0}$, for $\mu=120$GeV. $\;$
(b) $m_{h^0}$, for $\mu=300$GeV. $\;$
(c) $m_{A^0}$, for $\mu=120$GeV. $\;$
(d) $m_{A^0}$, for $\mu=300$GeV. $\;$
Values in the plots are in GeV. In all plots
$\tan\beta=50$, $A=0$, and $m_F=m_{F_L}=m_{F_R}$.
}
\label{f:mh0_and_mA0}
\end{figure}

In figure \ref{f:mh0_and_mA0} we plot the spectrum of the two neutral
Higgs bosons $h^0$ and $A^0$. 
For low $M_{1/2}$ their masses are degenerate while
for higher values of $M_{1/2}$ the pseudoscalar Higgs becomes degenerate
with the heavier of the two $CP$ even Higgs states. Our analysis shows
that the mass of
the lighter $CP$ even state is preferred to be in the range $112$--$117$GeV
for soft SUSY masses below $1\,$TeV. 
The pseudoscalar mass is quite sensitive to the
magnitude of the $D$ terms and, as was explained earlier, it was 
mildly pushed towards lower values as an additional condition 
on top of the REWSB conditions.

\begin{figure}[p]
\epsfysize=6.5truein
\epsffile{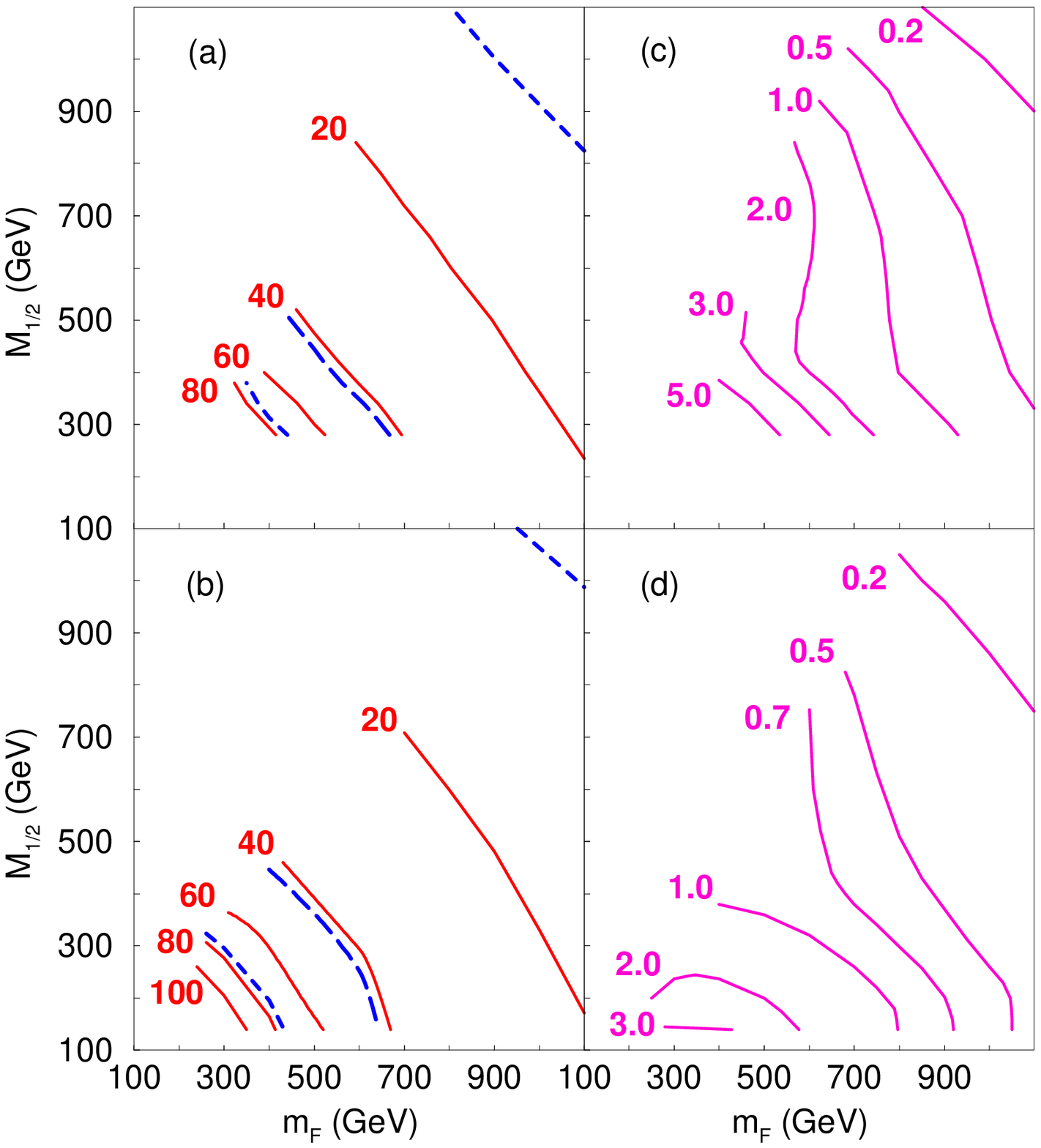}
\caption{
Contour lines of $\:\delta a_\mu({SUSY})\times 10^{10}$ 
and $\:BR$(\taumugamma)$\times 10^{6}$,
for two different values of $\mu$.
(a) $\delta a_\mu({SUSY})\times 10^{10}$, for $\mu=120$GeV. $\;$
(b) $\delta a_\mu({SUSY})\times 10^{10}$, for $\mu=300$GeV. $\;$
(c) $BR$(\taumugamma)$\times 10^{6}$,      for $\mu=120$GeV. $\;$
(d) $BR$(\taumugamma)$\times 10^{6}$,      for $\mu=300$GeV. $\;$
In (a) and (b) the long-dashed curve marks the central value for
$a_\mu$ not accounted for by the Standard Model, while the short-dashed
curves mark the 2$\sigma$ limits of this quantity. 
The experimental upper limit on $\:BR$(\taumugamma) 
is $1.1\times 10^{6}$ \cite{PDG}.
In all plots
$\tan\beta=50$, $A=0$, and $m_F=m_{F_L}=m_{F_R}$.
}
\label{f:amu_and_tmg}
\end{figure}

Figure \ref{f:amu_and_tmg} represents the main results of this study.
It shows that the constraints from the recent BNL experiment are 
consistent with all other constraints imposed on the model.
In fact, as shown in plots (a) and (b) the BNL $2\sigma$ region 
practically overlaps with the 
portion of the $(m_F,M_{1/2})$ plane below $1\,$TeV 
allowed by the direct sparticle searches. 
As promised in the text after Eq.\ref{Yuksq} we also focused
on the contribution to $a_\mu$ from the 22 entry in the slepton
matrix in (\ref{delta})
generated by the large 23 entry in (\ref{Yuk}). In our numerical
analysis the $\chi^2$ minimization procedure was extended to maximise
this contribution. Nevertheless the maximum enhancement we found
was on the level of 6\%. 

The large 23 entry in (\ref{Yuk})
makes an important contribution to the lepton flavour violating decay
\taumugamma. Plots (c) and (d) present the contour lines obtained
in the same analysis. The computed values should be compared to the 
experimental
upper limit $BR($\taumugamma$) < 1.1\times 10^{-6}$ at $90$\% C.L..
These predictions are quite robust. In order to reduce this branching
ratio below the experimental limit over the entire plane we
found we had to vary all initial parameters to rather extreme values, 
including lowering 
$\tan\beta$ as much as by 10 and increasing the trilinear parameter
$A$ into the TeV range.

In conclusion, we have discussed the lepton sector of
a realistic string-inspired model
 based on the Pati-Salam $SU(4)\times SU(2)_L \times SU(2)_R$ 
 gauge group supplemented by a $U(1)$ family symmetry. 
 The model involves third family Yukawa unification, predicts
 large $\tan \beta\sim 50$, and describes all fermion masses and
mixing angles, including approximate bi-maximal mixing in the neutrino sector.
 In particular atmospheric neutrino mixing is achieved via a large
 23 entry in the neutrino Yukawa matrix which we have shown to have important
phenomenological effects.
 We find that the recent
 BNL result on the muon ($g-2$) can be easily accommodated
in a large portion of the SUSY parameter space of the model.
Over this region of parameter space
the model predicts a
CP-even Higgs boson mass near 115 GeV,
and a rate for \taumugamma 
which is close to the current experimental limit.
We find it encouraging that 
all of these phenomenological features 
can be simultaneously
accomodated within a simple string-inspired model
such as the one considered in this study.

\vskip 0.1in
\noindent
 {\large {\bf Acknowledgments}}\\
T.B. would like to thank 
R.~Derm\'{\i}\v{s}ek 
for his help regarding the numerical procedure
used in this analysis.
S.K. thanks PPARC for a Senior Fellowship.

\newpage


\newpage

\end{document}